\documentclass[12pt,letterpaper]{article}
\usepackage{amsmath,amssymb,array,calc,rotating,epsfig,psfrag,amscd, cite}

\makeatletter
\renewcommand\section{\@startsection {section}{1}{\z@}%
                                   {-3.5ex \@plus -1ex \@minus -.2ex}
                                   {2.3ex \@plus.2ex}%
                                   {\normalfont\large\bfseries}}
\renewcommand\subsection{\@startsection{subsection}{2}{\z@}%
                                     {-3.25ex\@plus -1ex \@minus -.2ex}%
                                     {1.5ex \@plus .2ex}%
                                     {\normalfont\bfseries}}
\makeatother

\let\non\nonumber

\let\s=\sigma

\let\S=\Sigma

\newcommand{\bea}{\begin{eqnarray}}
\newcommand{\eea}{\end{eqnarray}}
\newcommand{\be}{\begin{equation}}
\newcommand{\ee}{\end{equation}}

\newcommand{\m}{\mu}

\newcommand{\p}{\partial}

\newcommand{\C}[1]{$(\ref{#1})$}

\def\IZ{\relax\ifmmode\mathchoice
{\hbox{\cmss Z\kern-.4em Z}}{\hbox{\cmss Z\kern-.4em Z}}
{\lower.9pt\hbox{\cmsss Z\kern-.4em Z}} {\lower1.2pt\hbox{\cmsss
Z\kern-.4em Z}}\else{\cmss Z\kern-.4em Z}\fi}
\def\IR{\relax{\rm I\kern-.18em R}}

\def\one{{\hbox{ 1\kern-.8mm l}}}

\newlength{\bredde}
\def\slash#1{\settowidth{\bredde}{$#1$}\ifmmode\,\raisebox{.15ex}{/}
\hspace*{-\bredde} #1\else$\,\raisebox{.15ex}{/}\hspace*{-\bredde}
#1$\fi}

\newsavebox{\zzzbar}
\sbox{\zzzbar}
  {\setlength{\unitlength}{0.9em}
  \begin{picture}(0.6,0.7)
  \thinlines
  \put(0,0){\line(1,0){0.6}}
  \put(0,0.75){\line(1,0){0.575}}
  \multiput(0,0)(0.0125,0.025){30}{\rule{0.3pt}{0.3pt}}
  \multiput(0.2,0)(0.0125,0.025){30}{\rule{0.3pt}{0.3pt}}
  \put(0,0.75){\line(0,-1){0.15}}
  \put(0.015,0.75){\line(0,-1){0.1}}
  \put(0.03,0.75){\line(0,-1){0.075}}
  \put(0.045,0.75){\line(0,-1){0.05}}
  \put(0.05,0.75){\line(0,-1){0.025}}
  \put(0.6,0){\line(0,1){0.15}}
  \put(0.585,0){\line(0,1){0.1}}
  \put(0.57,0){\line(0,1){0.075}}
  \put(0.555,0){\line(0,1){0.05}}
  \put(0.55,0){\line(0,1){0.025}}
  \end{picture}}

\newcommand{\ena}{\end{eqnarray}}
\newcommand{\beqa}{\begin{eqnarray}}
\newcommand{\eeqa}{\end{eqnarray}}

\def\m{\mu}

\def\s{\sigma}

\def\S{\Sigma}

\begin{document}
\begin{titlepage}

\begin{center}



\vskip 2 cm
{\Large \bf Poisson equation for the three loop ladder diagram in string theory at genus one}\\
\vskip 1.25 cm { Anirban Basu\footnote{email address:
    anirbanbasu@hri.res.in} } \\
{\vskip 0.5cm Harish--Chandra Research Institute, Chhatnag Road, Jhusi,\\
Allahabad 211019, India\\}

\end{center}

\vskip 2 cm

\begin{abstract}
\baselineskip=18pt

The three loop ladder diagram is a graph with six links and four cubic vertices that contributes to the $D^{12} \mathcal{R}^4$ amplitude at genus one in type II string theory. The vertices represent the insertion points of vertex operators on the toroidal worldsheet and the links represent scalar Green functions connecting them. By using the properties of the Green function and manipulating the various expressions, we obtain a modular invariant Poisson equation satisfied by this diagram, with source terms involving one, two and three loop diagrams. Unlike the source terms in the Poisson equations for diagrams at lower orders in the momentum expansion or the Mercedes diagram, a particular source term involves a five point function containing a holomorphic and a antiholomorphic worldsheet derivative acting on different Green functions. We also obtain simple equalities between topologically distinct diagrams, and consider some elementary examples.

\end{abstract}

\end{titlepage}


\section{Introduction}

Scattering amplitudes in type II string theory in ten dimensions lead to terms in the effective action which are analytic as well as non--analytic in the external momenta, in the low momentum expansion. The coefficients of these various contributions at genus $g$ are of the form $c_g e^{-2(1-g)\phi}$, where $\phi$ is the dilaton and $c_g$ are constants. At every genus, these constants are obtained from the explicit calculation of worldsheet correlation functions. At genus one, these involve inserting the various vertex operators on the toroidal worldsheet, computing the correlators taking care of various zero modes, integrating over the positions of the integrated vertex operators and finally integrating over the fundamental domain of $SL(2,\mathbb{Z})$ parametrized by the complex structure of the torus. Thus this boils down to calculating integrals of the form
\be \int_{\mathcal{F}} \frac{d^2\tau}{\tau_2^2} f(\tau,\bar\tau;s_{ij})\ee       
where $\mathcal{F}$ is the fundamental domain of $SL(2,\mathbb{Z})$ given by
\be \label{F}\mathcal{F} = \Big\{ -\frac{1}{2} \leq \tau_1 \leq \frac{1}{2}, \vert \tau \vert \geq 1 \Big\}\ee
and $f(\tau,\bar\tau;s_{ij})$ is modular invariant which depends on the Mandelstam variables $s_{ij} = -(k_i + k_j)^2$, where $k_i$ is the momentum of the state $i$. Such expressions contain terms analytic as well as non--analytic in $s_{ij}$. The non--analytic terms arise from the boundary of moduli space. 

To obtain both these contributions, the region in \C{F} is split into two regions: (i) $\tau_2 \leq L$, and (ii) $\tau_2 > L$, and each region is analyzed separately. Clearly the $L$ dependence cancels in the final answer, which is taken very large in the analysis. Thus the region (i) involves contributions from the bulk of moduli space, and here $f(\tau,\bar\tau;s_{ij})$ is expanded around $s_{ij}=0$ to yield various analytic terms in the amplitude. Region (ii) as $L \rightarrow \infty$ yields contributions from the boundary of moduli space where the integral is performed by using the asymptotic expansion of $f(\tau,\bar\tau;s_{ij})$. Note that this part of the analysis is non--perturbative in $s_{ij}$.        

Apart from determining these coefficients, these calculations assume significance in the context of U--duality of the toroidally compactified type II theory. The U--duality covariant moduli dependent coefficients of the various terms in the effective action admit a weak coupling expansion, and the result for the genus one amplitude in the appropriate dimension must match the worldsheet calculation. Thus the perturbative calculations are intricately tied to the non--perturbative duality symmetry of the theory. In particular, for certain BPS interactions, one obtains exact expressions~\cite{Green:1997tv,Green:1997as,Kiritsis:1997em,Green:1998by,Green:1999pu,Green:2005ba,Berkovits:2006vc,Basu:2007ru,Basu:2007ck,Basu:2008cf,Green:2010wi,Green:2010kv,Basu:2011he,Basu:2014hsa,Bossard:2014aea,Pioline:2015yea,Bossard:2015uga,Bossard:2015oxa,Bossard:2015foa,Basu:2016kon} which match expectations from string perturbation theory~\cite{Green:1999pv,D'Hoker:2005jc,D'Hoker:2005ht,Green:2008uj,D'Hoker:2013eea,Gomez:2013sla,D'Hoker:2014gfa,Basu:2015dqa,Pioline:2015nfa,Florakis:2016boz}.

Our focus is on the analytic part of the four graviton amplitude at genus one in ten dimensions, the low momentum expansion of which yields the genus one contribution to the $D^{2k} \mathcal{R}^4$ interactions. At every order in the momentum expansion, this yields integrals involving certain $SL(2,\mathbb{Z})$ invariant integrands. Hence each integrand is an $SL(2,\mathbb{Z})$ invariant non--holomorphic modular form. The structure of each integrand is determined by the topology of the various graphs on the toroidal worldsheet. These graphs arise from the low momentum expansion of the string amplitude which involves bringing down powers of the scalar Green function from the Koba--Nielsen factor. Thus the  
 links along these graphs are the Green functions that connect the vertices, while the vertices of the graphs are the positions of insertions of the vertex operators on the toroidal worldsheet. This leads to graphs with various topologies arising from various ways of connecting the vertices. Thus an analysis of these integrands is important in order to calculate string amplitudes. They also provide a rich arena for studying modular forms in mathematics which is interesting in its own right.

At leading orders in the momentum expansion, the various integrands that arise upto the $D^{10} \mathcal{R}^4$ interaction have been shown to obey a rich structure of Poisson equations by analyzing their detailed properties~\cite{D'Hoker:2015foa,D'Hoker:2015zfa,D'Hoker:2015qmf,D'Hoker:2016jac} and their contributions to the effective action have been evaluated~\cite{D'Hoker:2015foa,Basu:2016fpd}. Now among the various terms that arise for the $D^{12}\mathcal{R}^4$ interaction, there are only two contributions which involve graphs with cubic vertices that did not arise at lower orders in the momentum expansion. These are the Mercedes and three loop ladder diagrams given in figures 1 and 2 respectively. Both of them have four vertices and six links connecting them.     

\begin{figure}[ht]
\begin{center}
\[
\mbox{\begin{picture}(90,90)(0,0)
\includegraphics[scale=.45]{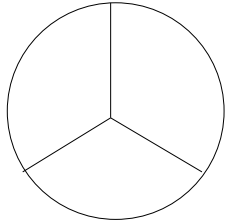}
\end{picture}}
\]
\caption{The Mercedes diagram}
\end{center}
\end{figure}

\begin{figure}[ht]
\begin{center}
\[
\mbox{\begin{picture}(110,60)(0,0)
\includegraphics[scale=.55]{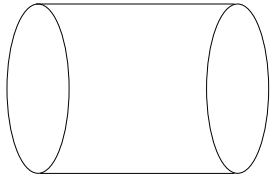}
\end{picture}}
\]
\caption{The three loop ladder diagram $\mathcal{L}$}
\end{center}
\end{figure}

Now the Mercedes diagram satisfies a Poisson equation~\cite{Basu:2015ayg}, which was obtained by expressing it directly in terms of integrated Green functions on the torus and using its various properties, which is very different from the line of analysis in~\cite{D'Hoker:2015zfa,D'Hoker:2016jac}.  Using the same techniques we obtain the Poisson equation satisfied by the three loop ladder diagram $\mathcal{L}$. The source terms in this equation involve several one, two and three loop diagrams with specific topologies. However, compared to the source terms in the Poisson equations for the diagrams upto the $D^{10}\mathcal{R}^4$ interaction as well as the Mercedes diagram, there is a difference. While all the other source terms in the various equations involve diagrams containing only the Green functions, the Poisson equation for the three loop ladder diagram has a source term that involves one holomorphic and one antiholomorphic worldsheet derivative acting on different Green functions in that diagram. The derivatives in this diagram, which is a five point function, cannot be removed. Like the other diagrams in the source terms for the various Poisson equations which either arise in the expressions for multi--graviton amplitudes or are expected to do so, this term is also expected to arise from such an amplitude. This turns out to be the case, and this is indeed a term that arises in the expression for the five graviton amplitude~\cite{Green:2013bza}. Thus we see how the structure of various multi--graviton amplitudes get intricately related.                      

We begin with a brief review of the four graviton amplitude at genus one in type II string theory. We then derive the Poisson equation for the three loop ladder diagram using various properties of the Green functions. Manipulations of these Green functions lead to several non--trivial identities among diagrams involving different topologies. We deduce some identities of this type in the final section, and look at some simple examples.   

\section{The four graviton amplitude in type II string theory at genus one}

The four graviton amplitude at genus one in type II superstring theory in ten dimensions is given by
\be \mathcal{A}_4 = 2\pi \mathcal{I}(s,t,u) \mathcal{R}^4, \ee
where
\be \label{oneloop}\mathcal{I} (s,t,u) = \int_{\mathcal{F}} \frac{d^2\tau}{\tau_2^2} F(s,t,u;\tau,\bar\tau).\ee
Here the Mandelstam variables $s, t, u$ satisfy the on--shell condition
\be s+t+u=0.\ee 
Also we have defined the measure $d^2 \tau = d\tau_1 d\tau_2$. The factor $F(s,t,u;\tau,\bar\tau)$ which encodes the worldsheet moduli and momentum dependence is given by
\be \label{D}F(s,t,u;\tau,\bar\tau) = \prod_{i=1}^4 \int_\S  \frac{d^2 z^{(i)}}{\tau_2} e^{\mathcal{D}},\ee
where $z^{(i)}$ $(i=1,2,3,4)$ are the positions of insertions of the four vertex operators on the toroidal worldsheet $\S$. Thus $d^2 z^{(i)} = d({\rm Re} z^{(i)}) d({\rm Im}z^{(i)})$, where
\be -\frac{1}{2} \leq {\rm Re} z^{(i)} \leq \frac{1}{2}, \quad 0 \leq {\rm Im} z^{(i)}\leq \tau_2 \ee
for all $i$. In \C{D}, the expression for $\mathcal{D}$ is given by
\be \label{defD}4\mathcal{D} = \alpha' s (\hat{G}_{12} + \hat{G}_{34})+\alpha' t (\hat{G}_{14} + \hat{G}_{23})+ \alpha' u (\hat{G}_{13} +\hat{G}_{24}),\ee
where $\hat{G}_{ij}$ is the scalar Green function on the torus with complex structure $\tau$ between points $z^{(i)}$ and $z^{(j)}$, and so
\be \hat{G}_{ij} \equiv \hat{G}(z^{(i)} - z^{(j)};\tau).\ee
Its explicit expression is given by~\cite{Green:1999pv,Green:2008uj}
\bea \label{prop}\hat{G}(z;\tau) &=& -{\rm ln} \Big\vert \frac{\theta_1 (z\vert\tau)}{\theta_1'(0\vert\tau)} \Big\vert^2 + \frac{2\pi({\rm Im}z)^2}{\tau_2} \non \\ &=& \frac{1}{\pi} \sum_{(m,n)\neq(0,0)} \frac{\tau_2}{\vert m\tau+n\vert^2} e^{\pi[\bar{z}(m\tau+n)-z(m\bar\tau+n)]/\tau_2} + 2{\rm ln} \vert \sqrt{2\pi} \eta(\tau)\vert^2.\eea
Note that the $z$ independent zero mode given by the second term in the second line of \C{prop} cancels in the whole amplitude, which follows from the expression for $\mathcal{D}$ in \C{defD} on using $s+t+u=0$. Thus in the expression for $\mathcal{D}$ we simply replace $\hat{G}(z;\tau)$ by $G(z;\tau)$ where
\be \label{Green}G(z;\tau) = \frac{1}{\pi} \sum_{(m,n)\neq(0,0)} \frac{\tau_2}{\vert m\tau+n\vert^2} e^{\pi[\bar{z}(m\tau+n)-z(m\bar\tau+n)]/\tau_2},\ee
where $G(z;\tau)$ is modular invariant, and single valued. Thus
\be \label{sv}G(z;\tau) = G(z+1;\tau) = G(z+\tau;\tau).\ee
It is the Green function $G(z;\tau)$ that arises in the expressions for higher point graviton amplitudes as well which follows from the modular invariance of the amplitudes. 

As mentioned in the introduction, in \C{oneloop}, $\mathcal{F}$ is split into
\be \mathcal{F} = \mathcal{F}_L +\mathcal{R}_L,\ee
where $\mathcal{F}_L$ is defined for $\tau_2 \leq L$, and $\mathcal{R}_L$ is defined for $\tau_2 > L$. Thus the analytic part of the amplitude is given by
\be \mathcal{I}_{an} (s,t,u) =   \sum_{n=0}^\infty\int_{\mathcal{F}_L} \frac{d^2\tau}{\tau_2^2}\prod_{i=1}^4 \int_\S  \frac{d^2 z^{(i)}}{\tau_2} \cdot \frac{\mathcal{D}^n}{n!}.\ee
Hence performing an $\alpha'$ expansion gives us
\be \mathcal{I}_{an} (s,t,u)= \sum_{p,q} \s_2^p \s_3^q J^{p,q},\ee
where
\be J^{p,q} = \int_{\mathcal{F}_L} \frac{d^2\tau}{\tau_2^2} j^{p,q} (\tau,\bar\tau),\ee
and
\be \s_2 = \alpha'^2 (s^2 +t^2 +u^2), \quad \s_3 = \alpha'^3 (s^3 +t^3 +u^3).\ee
Thus we see that $j^{p,q}(\tau,\bar\tau)$ is obtained after integrating over the insertion points of the vertex operators and encodes the topologically distinct ways the scalar propagators are connected on the toroidal worldsheet.

The three loop ladder diagram first arises at order $D^{12} \mathcal{R}^4$ in the low momentum expansion. It is given by 
\be \label{l}\mathcal{L} = \int_\S  \frac{d^2 z^{(1)}}{\tau_2} \int_\S  \frac{d^2 z^{(2)}}{\tau_2} \int_\S  \frac{d^2 z^{(3)}}{\tau_2}  \int_\S \frac{d^2 z^{(4)}}{\tau_2} G_{12}^2  G_{34}^2 G_{13} G_{24} \ee
as given in figure 2.

The contribution of the ladder diagram\footnote{We shall refer to the three loop ladder diagram simply as the ladder diagram from now onwards for brevity.} to the $D^{12}\mathcal{R}^4$ interaction is given by (see equations (C.5) ad (C.6) in~\cite{Green:2008uj})
\bea \label{cont} j^{(3,0)} = -\frac{\mathcal{L}}{16} ,\quad j^{(0,2)} = \frac{\mathcal{L}}{4}.\eea 
In obtaining these numerical factors, we have used the equality \C{6} to include the contributions from both these diagrams.  

\section{The Poisson equation for the three loop ladder diagram}

We want to derive the Poisson equation the ladder diagram \C{l} satisfies. To do so, we make use of the various properties satisfied by the Green function on the torus (see~\cite{D'Hoker:2015foa,Basu:2015ayg} for various details). We find it very useful to use the relations satisfied by them under the variation of the Beltrami differential $\m$. We have that
\be \label{onevar} \p_\mu G(z_1,z_2) = -\frac{1}{\pi} \int_\S d^2 z \p_z G(z,z_1) \p_z G(z,z_2),\ee  
and
\be \bar\p_{\m}\p_\m G(z_1,z_2) =0.\ee
Also the $SL(2,\mathbb{Z})$ invariant Laplacian is defined by
\be \label{beltrami}\Delta = 4\tau_2^2\frac{\p^2}{\p\tau\p\bar\tau} = \bar\p_{\m} \p_\m.\ee

The Green function satisfies the equations
\bea \label{eigen}\bar{\p}_w\p_z G(z,w) = \pi \delta^2 (z-w) - \frac{\pi}{\tau_2}, \non \\ \bar{\p}_z\p_z G(z,w) = -\pi \delta^2 (z-w) + \frac{\pi}{\tau_2} \eea
which is repeatedly used in our analysis. 

In the various manipulations, we often obtain expressions involving $\p_z G(z,w)$ where $z$ is integrated over $\S$. We then integrate by parts without picking up boundary contributions on $\S$ as $G(z,w)$ is single valued. Also we readily use $\p_z G(z,w) = -\p_w G(z,w)$ using the translational invariance of the Green function. Finally, we have that
\be \int_\S d^2 z G(z,w)=0\ee
which easily follows from \C{Green}.

In the various expressions, for brevity we write
\be \int_\S d^2 z \int_\S d^2 w \ldots \equiv \int_{zw\ldots}.\ee

Now from \C{beltrami} we have that 
\be \label{L}\Delta \mathcal{L} = \p_\mu \bar\p_{\mu} \mathcal{L} = 4\mathcal{L}_1 + 8(\mathcal{L}_2 + c.c.) + 2\mathcal{L}_3 + 8\mathcal{L}_4,\ee 
where $\mathcal{L}_1$, $\mathcal{L}_2$, $\mathcal{L}_3$ and $\mathcal{L}_4$ are defined by
\bea \label{L1234} \mathcal{L}_1 &=& \frac{1}{\tau_2^4} \int_{1234} \p_\mu G_{12} \bar\p_{\mu} G_{12} G_{34}^2 G_{13} G_{24}, \non \\ \mathcal{L}_2 &=& \frac{1}{\tau_2^4} \int_{1234} G_{12} \p_{\mu} G_{12} G_{34}^2 \bar\p_{\mu} G_{13} G_{24} , \non \\ \mathcal{L}_3 &=&\frac{1}{\tau_2^4} \int_{1234}  G_{12}^2 G_{34}^2 \p_{\mu}G_{13} \bar\p_{\mu}G_{24}, \non \\ \mathcal{L}_4 &=&\frac{1}{\tau_2^4} \int_{1234}  G_{12} \p_\m G_{12} G_{34} \bar\p_{\mu}G_{34} G_{13} G_{24}.\eea

These four topologically distinct contributions are given in figure 3. In these diagrams, $\mu$ along a link stands for $\p_\mu$, while $\bar\mu$ stands for $\bar\p_{\mu}$.

\begin{figure}[ht]
\begin{center}
\[
\mbox{\begin{picture}(245,190)(0,0)
\includegraphics[scale=.6]{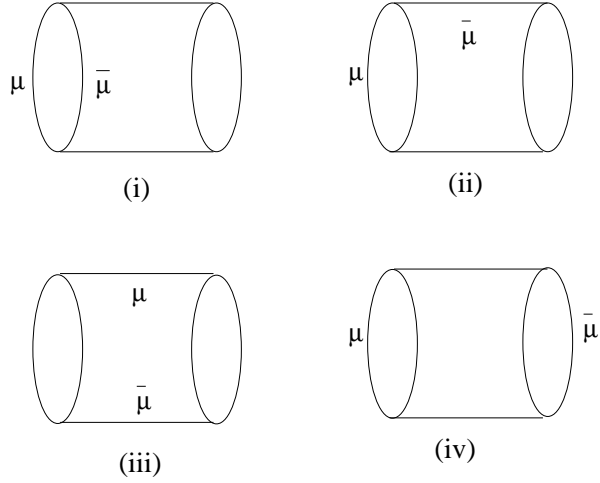}
\end{picture}}
\]
\caption{The diagrams (i) $\mathcal{L}_1$, (ii) $\mathcal{L}_2$, (iii) $\mathcal{L}_3$ and (iv) $\mathcal{L}_4$}
\end{center}
\end{figure}

In our analysis below, it shall be very convenient to depict the various relations using diagrams. The notations for holomorphic and antiholomorphic derivatives acting on the Green function are given in figure 4. From the structure of \C{Green} it follows that one particle reducible diagrams vanish and hence we ignore them. Also diagrams of the form given in figure 5 for any $A$, or its complex conjugate vanish as they are total derivatives, and we ignore them as well.

\begin{figure}[ht]
\begin{center}
\[
\mbox{\begin{picture}(240,50)(0,0)
\includegraphics[scale=.75]{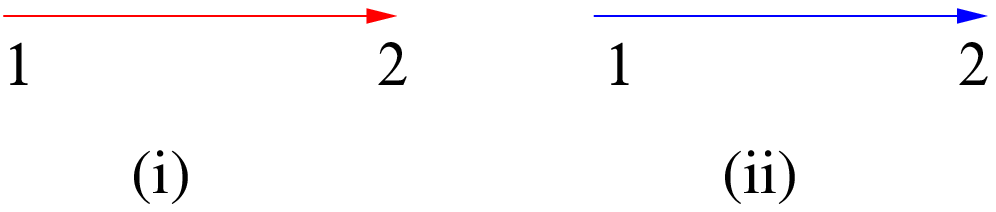}
\end{picture}}
\]
\caption{(i) $\p_2 G_{12} = -\p_1 G_{12}$, (ii) $\bar\p_2 G_{12} = -\bar\p_1 G_{12}$}
\end{center}
\end{figure}

\begin{figure}[ht]
\begin{center}
\[
\mbox{\begin{picture}(145,50)(0,0)
\includegraphics[scale=.55]{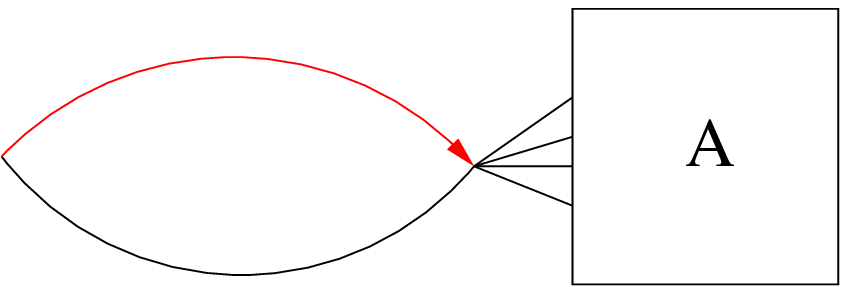}
\end{picture}}
\]
\caption{Vanishing contributions}
\end{center}
\end{figure}

We now consider each of the contributions from \C{L1234} that lead to \C{L} separately.

\subsection{The contribution from $\mathcal{L}_1$}

\begin{figure}[ht]
\begin{center}
\[
\mbox{\begin{picture}(395,70)(0,0)
\includegraphics[scale=.65]{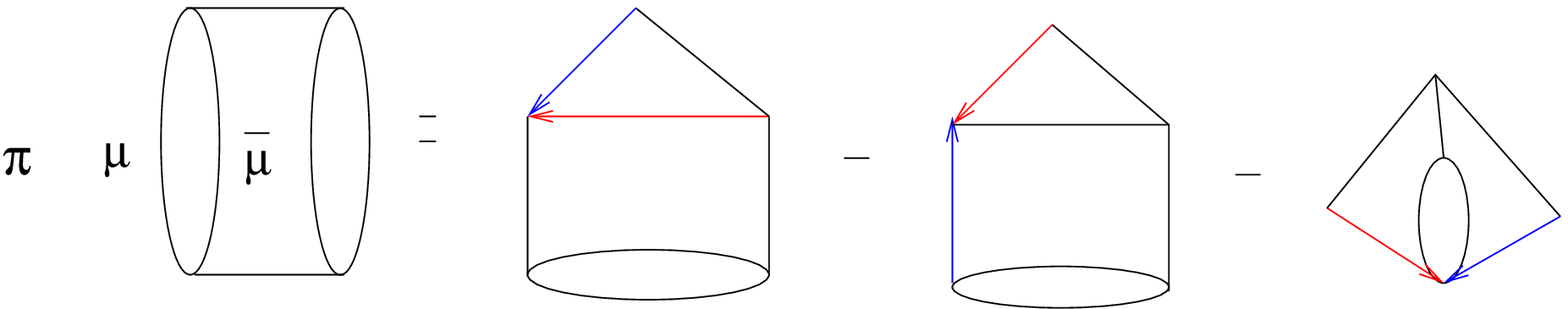}
\end{picture}}
\]
\caption{An intermediate equation for $\mathcal{L}_1$}
\end{center}
\end{figure}

We first consider the contribution coming from the diagram $\mathcal{L}_1$. From \C{onevar}, we see that it contains two $\p$s and two $\bar\p$s, which is the starting point of our analysis for $\mathcal{L}_2$, $\mathcal{L}_3$ and $\mathcal{L}_4$ as well. To begin with, we manipulate the four derivatives using \C{onevar} to integrate by parts and then use \C{eigen}, to express $\mathcal{L}_1$ in terms of diagrams containing only one $\p$ and one $\bar\p$. This is given by
\be \pi \mathcal{L}_1 = \mathcal{X}_1 - \mathcal{X}_2 - \mathcal{X}_3\ee
which is given in figure 6.

Here $\mathcal{X}_1$, $\mathcal{X}_2$ and $\mathcal{X}_3$ are five point functions defined by
\bea \mathcal{X}_1 &=& \frac{1}{\tau_2^4} \int_{12345} G_{12} \bar\p_2 G_{23} G_{34} G_{45} G_{51}^2 \p_2  G_{24}, \non \\ \mathcal{X}_2 &=& \frac{1}{\tau_2^4} \int_{12345} \bar\p_2 G_{12} \p_2 G_{23} G_{34} G_{45} G_{51}^2   G_{24} , \non \\ \mathcal{X}_3 &=&  \frac{1}{\tau_2^4}\int_{12345} \bar\p_1 G_{12} G_{23} G_{34} \p_1 G_{41} G_{15}^2 G_{35},\eea
which are given in figure 7. We first manipulate the expressions involving $\mathcal{X}_1$ and $\mathcal{X}_2$ to derive the relation
\be \label{one}\mathcal{X}_1 - \mathcal{X}_2 = -\pi \mathcal{F}_2 +\pi \mathcal{T}_2 \mathcal{T}_3 - \pi \mathcal{V}_1.\ee
We also get that
\be \label{two}\mathcal{X}_3 = -\frac{\pi}{2} \mathcal{F}_3 +\frac{\pi}{2} \mathcal{T}_2 \mathcal{F}_1 +\pi \mathcal{F}_2, \ee
where the various diagrams on the right hand sides of \C{one} and \C{two} are defined in the appendix. Again use is made of the relations \C{onevar} and \C{eigen} to obtain this equation, which is also true of the various equations later. Note that in this analysis as well as later, some graphs which arise at the end of the analysis factorize on using the explicit expression in \C{Green}.

\begin{figure}[ht]
\begin{center}
\[
\mbox{\begin{picture}(265,85)(0,0)
\includegraphics[scale=.65]{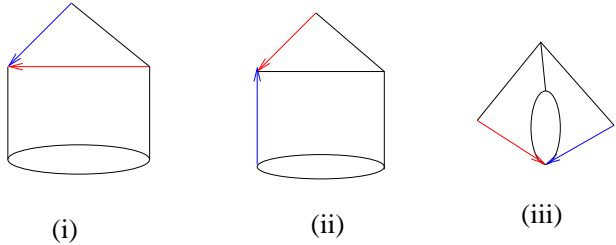}
\end{picture}}
\]
\caption{The diagrams (i) $\mathcal{X}_1$, (ii) $\mathcal{X}_2$ and (iii) $\mathcal{X}_3$}
\end{center}
\end{figure}

Thus finally we get that
\be \mathcal{L}_1 = -2\mathcal{F}_2 + \mathcal{T}_2 \mathcal{T}_3 - \mathcal{V}_1 +\frac{1}{2} \mathcal{F}_3 - \frac{1}{2} \mathcal{T}_2 \mathcal{F}_1\ee
which is given in figure 8. Thus $\mathcal{L}_1$ is expressed completely in terms of diagrams with no derivatives.  

\begin{figure}[ht]
\begin{center}
\[
\mbox{\begin{picture}(295,150)(0,0)
\includegraphics[scale=.6]{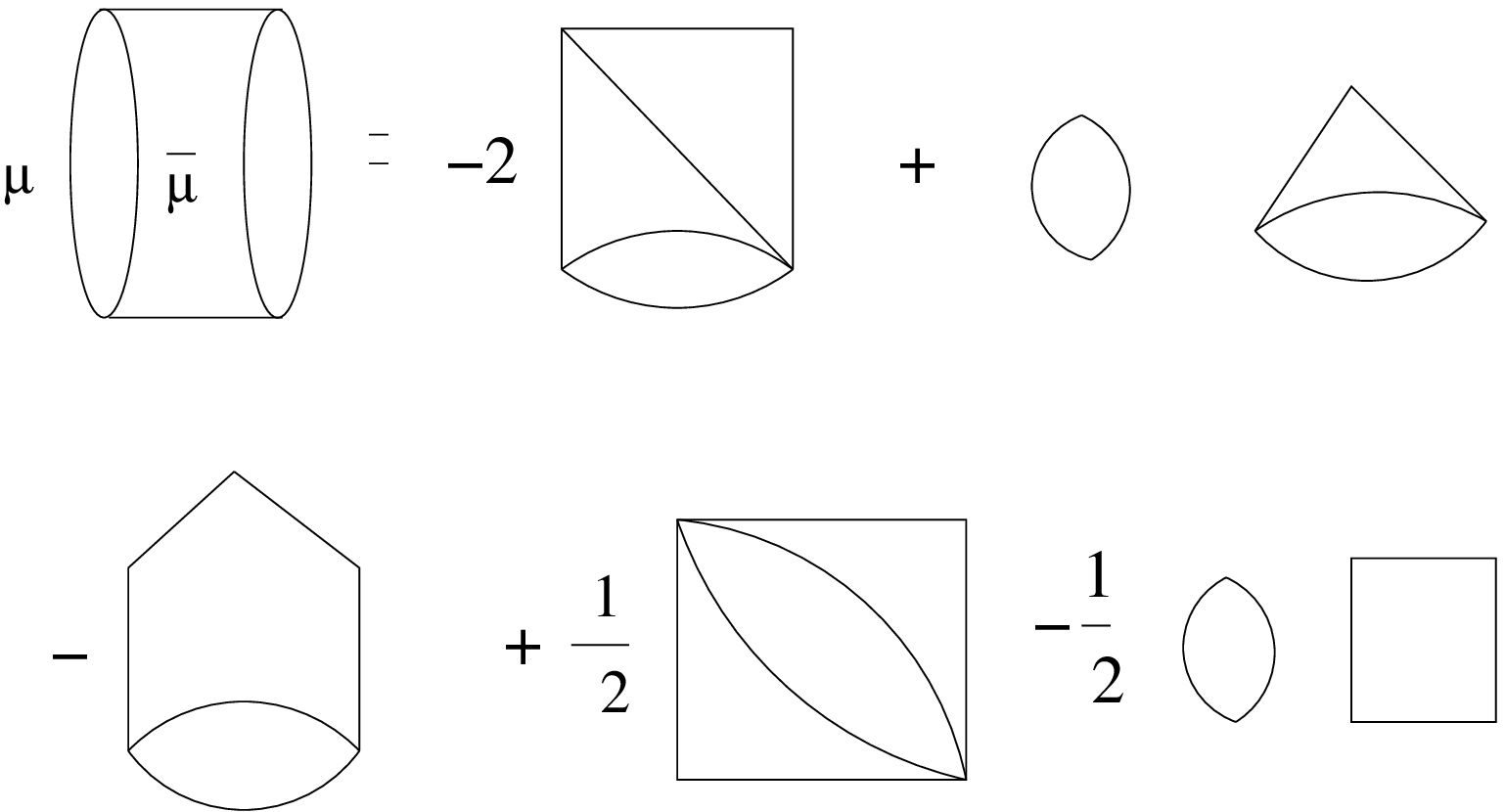}
\end{picture}}
\]
\caption{The equation for $\mathcal{L}_1$}
\end{center}
\end{figure}

\subsection{The contribution from $\mathcal{L}_2$}

We now consider the contribution from $\mathcal{L}_2$. Reducing the number of derivatives from four to zero and two, at an intermediate step we get that
\be \mathcal{L}_2 =  \mathcal{V}_1 - \frac{1}{2}\mathcal{L} +\frac{1}{\pi}\mathcal{Y} \ee
as given in figure 9. Here $\mathcal{Y}$ is a five point function defined by
\be \mathcal{Y} = \frac{1}{\tau_2^4} \int_{12345} \bar\p_2 G_{12} \p_2 G_{23} G_{34} G_{45} G_{51}^2 G_{24}\ee
as given in figure 10. 

\begin{figure}[ht]
\begin{center}
\[
\mbox{\begin{picture}(345,60)(0,0)
\includegraphics[scale=.6]{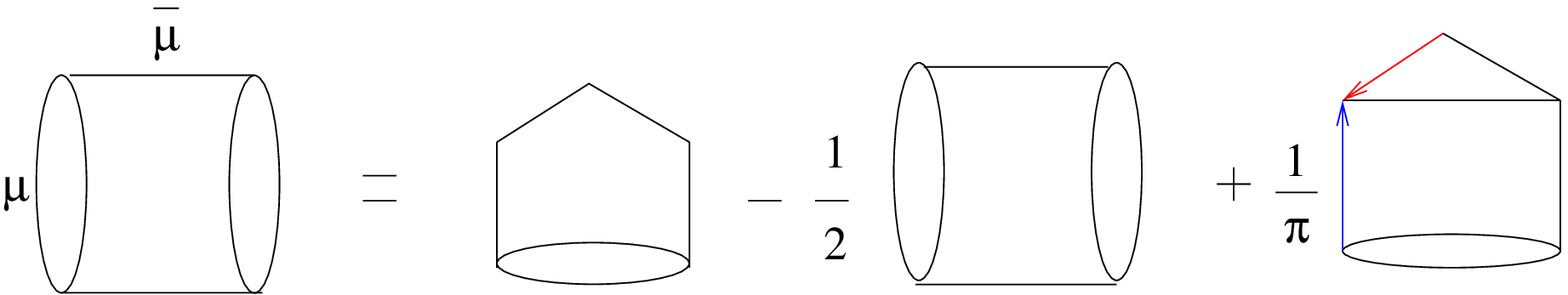}
\end{picture}}
\]
\caption{An intermediate equation for $\mathcal{L}_2$}
\end{center}
\end{figure}

\begin{figure}[ht]
\begin{center}
\[
\mbox{\begin{picture}(75,55)(0,0)
\includegraphics[scale=.6]{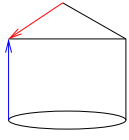}
\end{picture}}
\]
\caption{The diagram $\mathcal{Y}$}
\end{center}
\end{figure}

We now manipulate the sum of $\mathcal{Y}$ and its complex conjugate to obtain 
\be \label{Y}\mathcal{Y} + c.c. = \pi \mathcal{L} +\pi \mathcal{F}_2 - \pi \mathcal{T}_2 \mathcal{T}_3 + \pi \mathcal{V}_1, \ee
where $\mathcal{L}$ is the ladder diagram, and the others are defined in the appendix. Thus we have that
\be \mathcal{L}_2 + c.c. = 3 \mathcal{V}_1 + \mathcal{F}_2 - \mathcal{T}_2 \mathcal{T}_3.\ee
Hence the final expression for $\mathcal{L}_2$ involves no diagrams with any derivatives as given in figure 11. 

\begin{figure}[ht]
\begin{center}
\[
\mbox{\begin{picture}(325,50)(0,0)
\includegraphics[scale=.65]{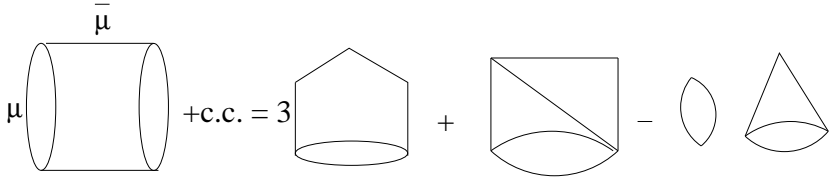}
\end{picture}}
\]
\caption{The equation for $\mathcal{L}_2$}
\end{center}
\end{figure}

\subsection{The contribution from $\mathcal{L}_3$}

We next consider the contribution from $\mathcal{L}_3$. Again proceeding along the lines of the above analysis, we get that
\be \mathcal{L}_3 = \mathcal{L}\ee
which is given in figure 12. 

\begin{figure}[ht]
\begin{center}
\[
\mbox{\begin{picture}(160,50)(0,0)
\includegraphics[scale=.65]{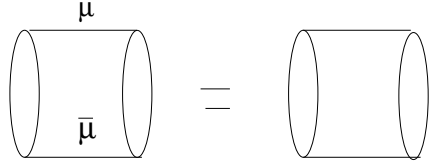}
\end{picture}}
\]
\caption{The equation for $\mathcal{L}_3$}
\end{center}
\end{figure}

\subsection{The contribution from $\mathcal{L}_4$}

Finally we consider the contribution from $\mathcal{L}_4$, which is the most involved one. Unlike the other cases, it is difficult to express $\mathcal{L}_4$ in terms of diagrams with only two derivatives starting directly from the one given in figure 3. Hence for this case we proceed differently.  

We consider an auxiliary diagram with six derivatives (three $\p$s and three $\bar\p$s) which trivially reduces to $\mathcal{L}_4$. On the other hand the this new diagram is such that it can be calculated in a different way such that it reduces to diagrams with at most two derivatives only. This diagram $\mathcal{Z}$ is given by
\be \mathcal{Z} = \frac{1}{\tau_2^4} \int_{12345} G_{12} \p_\mu G_{12} G_{34} \p_{\bar\m} G_{34} G_{24} \bar\p_1 G_{15} \p_3 G_{35}\ee
as given in figure 13. Now the integral over the position 5 can be done trivially leading to
\be \label{Def}\mathcal{Z} = \pi \mathcal{L}_4.\ee 

\begin{figure}[ht]
\begin{center}
\[
\mbox{\begin{picture}(100,55)(0,0)
\includegraphics[scale=.65]{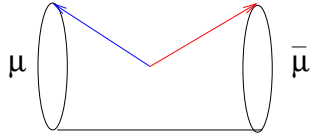}
\end{picture}}
\]
\caption{The diagram for $\mathcal{Z}$}
\end{center}
\end{figure}

This is obtained by simply putting the neighboring $\p$ and $\bar\p$ derivatives on the same line and using \C{eigen}. However we can also calculate $\mathcal{Z}$ differently by allowing the various derivatives to run along the various links. This gives us contributions to $\mathcal{Z}$ having diagrams involving at most two derivatives, and we have that
\be \label{E}\mathcal{Z} = \pi \mathcal{S}_1 +\frac{\pi}{4} \mathcal{L} -\pi \mathcal{V}_1 + \mathcal{V}_2 - (\mathcal{Z}_1 + c.c.) + (\mathcal{Z}_2 +c.c.),\ee
where $\mathcal{Z}_1$ and $\mathcal{Z}_2$ are defined by
\bea \mathcal{Z}_1= \frac{1}{\tau_2^4} \int_{12345} G_{12} \p_1 G_{12} G_{13} G_{34} G_{24} G_{45} \bar\p_3 G_{35}, \non \\ \mathcal{Z}_2 = \frac{1}{\tau_2^5} \int_{123456} G_{12} \p_3 G_{23} G_{34} G_{45} G_{51} G_{56} \bar\p_4 G_{46} ,\eea
which is given in figure 14. Hence the relation \C{E} is given in figure 15.

\begin{figure}[ht]
\begin{center}
\[
\mbox{\begin{picture}(140,75)(0,0)
\includegraphics[scale=.65]{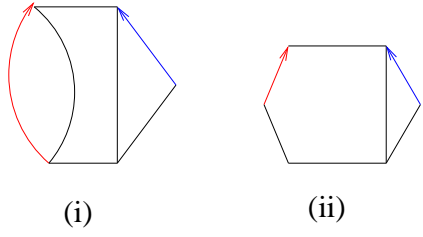}
\end{picture}}
\]
\caption{The diagrams for (i) $\mathcal{Z}_1$ and (ii) $\mathcal{Z}_2$}
\end{center}
\end{figure}

\begin{figure}[ht]
\begin{center}
\[
\mbox{\begin{picture}(330,205)(0,0)
\includegraphics[scale=.6]{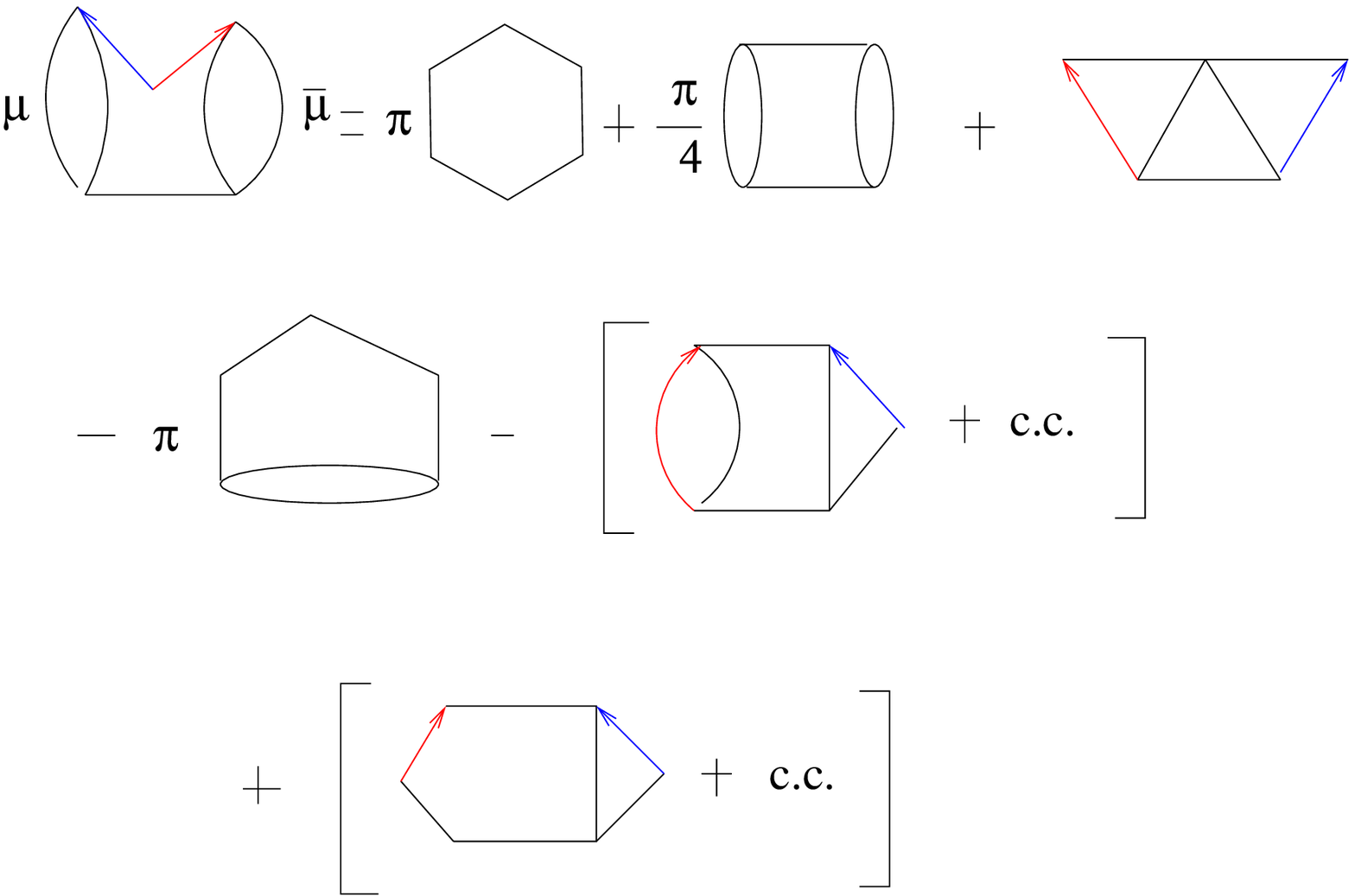}
\end{picture}}
\]
\caption{An intermediate equation for $\mathcal{Z}$}
\end{center}
\end{figure}

We now calculate the sum of $\mathcal{Z}_1$ and its complex conjugate leading to further simplifications. Using the relation 
\be \mathcal{Z}_1 = \frac{\mathcal{Y}^*}{2},\ee
from \C{Y} we get that
\be \mathcal{Z}_1 + c.c. = \frac{1}{2} (\mathcal{Y} + c.c.)= \frac{\pi}{2} (\mathcal{L} +\mathcal{F}_2 - \mathcal{T}_2 \mathcal{T}_3 + \mathcal{V}_1).\ee
 
Next we calculate the sum of $\mathcal{Z}_2$ and its complex conjugate which also simplifies. To do so, we start with the diagram $\tilde{\mathcal{Z}}$ defined by
\be \tilde{\mathcal{Z}} = \frac{1}{\tau_2^5} \int_{123456} G_{12} G_{23} G_{34} G_{45} G_{56} G_{61} \p_6 \bar\p_4  G_{46}\ee
which is given in figure 16. We evaluate it in two ways, one way using \C{eigen} for the $\p$ and $\bar\p$ on the same link, and the other by moving them around. This leads to

\begin{figure}[ht]
\begin{center}
\[
\mbox{\begin{picture}(100,105)(0,0)
\includegraphics[scale=.75]{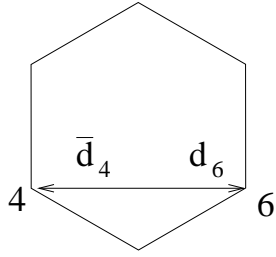}
\end{picture}}
\]
\caption{The diagram $\tilde{\mathcal{Z}}$}
\end{center}
\end{figure} 

\be \mathcal{Z}_2 + c.c. = \pi (\mathcal{V}_1 +\mathcal{V}_3 +\mathcal{S}_1 - \mathcal{T}_2 \mathcal{F}_1 ).\ee
Thus we see that the contributions involving $\mathcal{Z}_1$, $\mathcal{Z}_2$ and their complex conjugates all reduce to diagrams with no derivatives. Hence from \C{Def} and \C{E} we see that the only contribution to $\mathcal{L}_4$ which involves two derivatives is from the diagram $\mathcal{V}_2$, leading to
\be \mathcal{L}_4 = 2\mathcal{S}_1 +\mathcal{V}_3  - \frac{1}{2}\mathcal{V}_1 +\frac{1}{\pi}\mathcal{V}_2 -\frac{1}{4} \mathcal{L} - \frac{1}{2} \mathcal{F}_2 -\mathcal{T}_2 \mathcal{F}_1 +\frac{1}{2}\mathcal{T}_2 \mathcal{T}_3 ,\ee 
as given in figure 17. 

\begin{figure}[ht]
\begin{center}
\[
\mbox{\begin{picture}(220,155)(0,0)
\includegraphics[scale=.6]{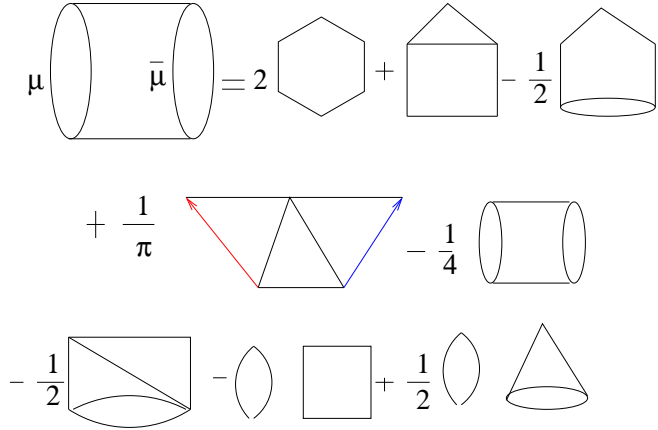}
\end{picture}}
\]
\caption{The equation for $\mathcal{L}_4$}
\end{center}
\end{figure} 

Now $\mathcal{V}_2$ is a contribution whose two derivatives cannot be removed. In fact, we obtain the relation
\be \mathcal{V}_2 = \mathcal{V}_4 +\pi \mathcal{V}_3 +\pi\mathcal{F}_2 - \pi \mathcal{T}_2 \mathcal{T}_3\ee 
from manipulating the diagram $\mathcal{V}_4$ which is defined in the appendix. This relation is given in figure 18.   

\begin{figure}[ht]
\begin{center}
\[
\mbox{\begin{picture}(260,115)(0,0)
\includegraphics[scale=.6]{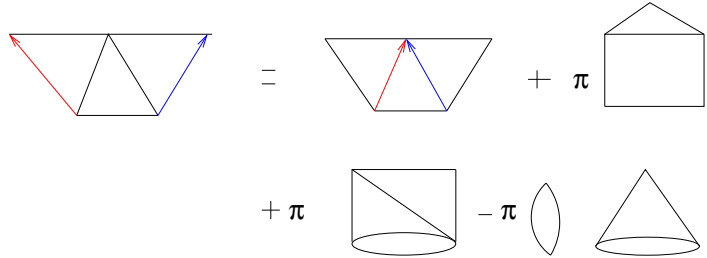}
\end{picture}}
\]
\caption{Relating $\mathcal{V}_2$ and $\mathcal{V}_4$}
\end{center}
\end{figure} 

Now $\mathcal{V}_4$ is precisely one of the 7 elements of the basis of diagrams involving two derivatives in the expression for the five graviton amplitude at genus one~\cite{Green:2013bza}. For these 7 diagrams, the derivatives cannot be eliminated. We see that $\mathcal{V}_2$ and $\mathcal{V}_4$ are equal upto terms where the derivatives can be eliminated. Thus like the other source terms in the various Poisson equations, this term has an interpretation as well in terms of multi--graviton amplitudes. Such kind of a term where the derivatives cannot be removed did not arise in the Poisson equations for the four graviton amplitude at lower orders in the momentum expansion, and it first shows up at this order. It was also not present in the Poisson equation for the Mercedes diagram.     

Thus adding the various contributions, we see that the ladder diagram satisfies the Poisson equation
\be \Delta \mathcal{L} = 16 \mathcal{S}_1 +\frac{8}{\pi} \mathcal{V}_2 +16 \mathcal{V}_1 +8\mathcal{V}_3 - 4\mathcal{F}_2 +2\mathcal{F}_3 -10 \mathcal{T}_2 \mathcal{F}_1,\ee
which is given in figure 19. 

\begin{figure}[ht]
\begin{center}
\[
\mbox{\begin{picture}(380,95)(0,0)
\includegraphics[scale=.7]{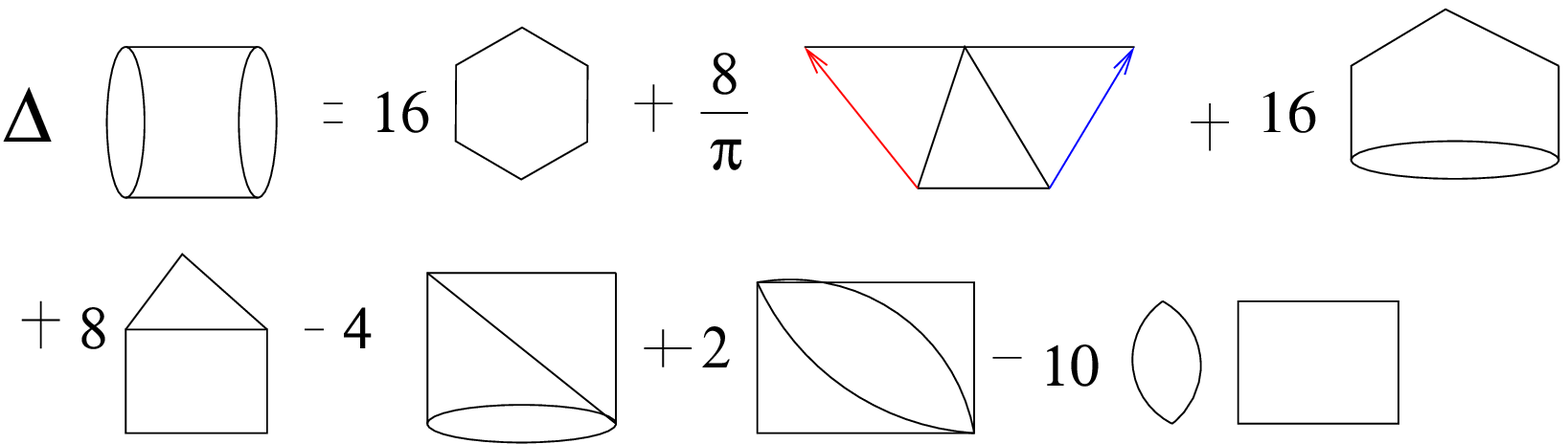}
\end{picture}}
\]
\caption{The Poisson equation for $\mathcal{L}$}
\end{center}
\end{figure} 

\section{Some relations among diagrams involving different topologies}

Any term in an amplitude at a fixed order in the low momentum expansion has several diagrams of different topologies that result from contracting the various Green functions (and the derivatives that can come associated with it) that arise in the expression for the genus one string amplitude. These are given by graphs with the same number of links, but can have different number of vertices. This is because the number of links is given by the number of Green functions which is fixed at a given order in the low momentum expansion, while the number of vertices is given by the number of vertex operator insertions points that are connected by the Green functions, and hence need not be the same for every diagram. We saw the appearance of several such diagrams in our analysis above. Now topologically inequivalent diagrams can be related. Several such relations have been given in~\cite{D'Hoker:2015foa} involving equations for such diagrams with a fixed number of links.   

In this section, we shall consider a simple diagrammatic treatment using the properties of Green functions to arrive at certain equalities involving topologically distinct diagrams. Proving these equalities directly from the expressions for the relevant diagrams is involved, however we shall see that they follow in a simple way from our analysis. We shall look at some elementary cases to illustrate the point, though the method can be used to derive other relations as well.    

Consider the equation that is given in figure 20, where the blobs $A, B$ and $C$ do not involve any $\p$ or $\bar\p$. It leads to an equality between three diagrams with three blobs which contain no $\p$ or $\bar\p$. While this relation is easily obtained, it leads to equalities between diagrams with different topologies for arbitrary choices of $A, B$ and $C$.

\begin{figure}[ht]
\begin{center}
\[
\mbox{\begin{picture}(380,60)(0,0)
\includegraphics[scale=.65]{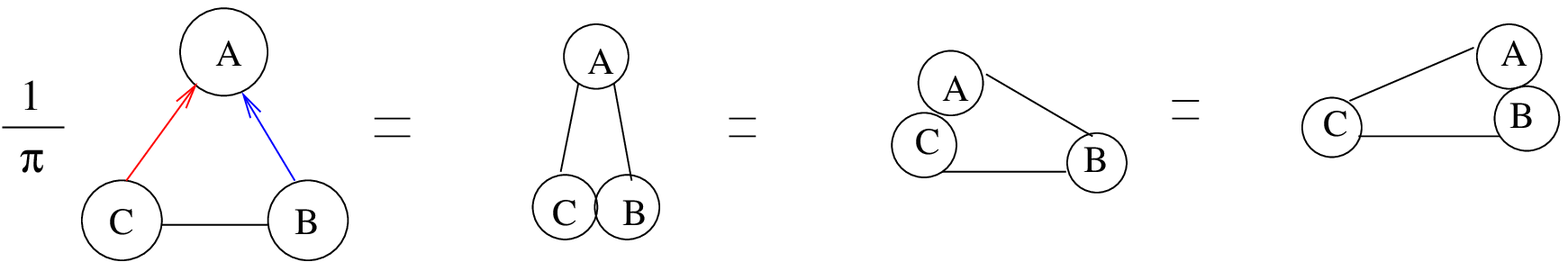}
\end{picture}}
\]
\caption{An equality among diagrams}
\end{center}
\end{figure} 

We now diagrammatically write down some consequences of this equality for specific choices of $A, B$ and $C$.

\begin{figure}[ht]
\begin{center}
\[
\mbox{\begin{picture}(160,60)(0,0)
\includegraphics[scale=.65]{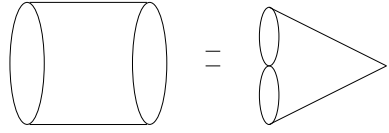}
\end{picture}}
\]
\caption{An equality between diagrams with 6 links}
\end{center}
\end{figure} 

For diagrams with 6 links, this leads to the relation in figure 21. Thus (we use the notation of~\cite{Green:2008uj}, where $\mathcal{L} = D_{1,2,1,2}$)
\be \label{6} \mathcal{L} = D_{1,1,2,2},\ee 
leading to an equality involving diagrams for the $D^{12} \mathcal{R}^4$ interaction. In fact, the leading terms in the large $\tau_2$ expansion indeed match, and are given by~\cite{Green:2008uj}
\be \pi^6\mathcal{L} = \pi^6 D_{1,1,2,2} = \frac{612}{691} \zeta(12)\tau_2^6 +\frac{8\pi}{3} \zeta(3)\zeta(8) \tau_2^3 - \pi \zeta(5)\zeta(6)\tau_2 +21 \zeta(3)^2\zeta(6) + O(\tau_2^{-1}).\ee

For diagrams with 7 links, we get the equations in figure 22. The equality involving four point functions equates diagrams relevant for the $D^{14} \mathcal{R}^4$ interaction, while that for five point functions equates diagrams relevant for the $D^{12}\mathcal{R}^5$ interaction.  

\begin{figure}[ht]
\begin{center}
\[
\mbox{\begin{picture}(160,120)(0,0)
\includegraphics[scale=.6]{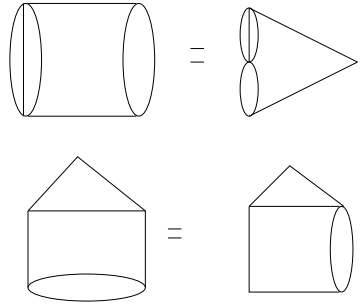}
\end{picture}}
\]
\caption{Equalities between diagrams with 7 links}
\end{center}
\end{figure}

\begin{figure}[ht]
\begin{center}
\[
\mbox{\begin{picture}(160,165)(0,0)
\includegraphics[scale=.7]{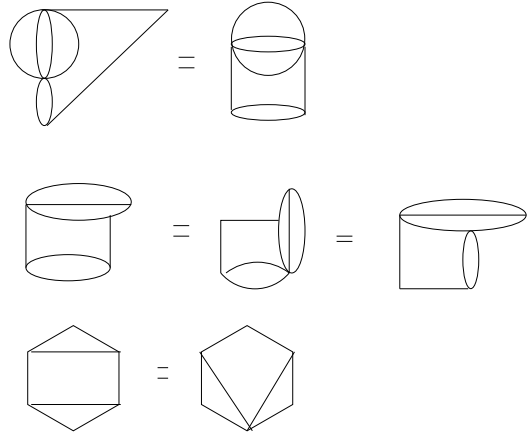}
\end{picture}}
\]
\caption{A few equalities between diagrams with 8 links}
\end{center}
\end{figure} 

One can continue to get non--trivial identities between various diagrams with higher number of links. 
We list equalities between only a few diagrams with 8 links in figure 23. The equality involving four point functions equates diagrams relevant for the $D^{16} \mathcal{R}^4$ interaction. The equalities involving the five and six point functions equate diagrams relevant for the $D^{14}\mathcal{R}^5$ and $D^{12} \mathcal{R}^6$ interactions respectively.

\appendix

\section{Relevant diagrams for the Poisson equation}

The Poisson equation for the ladder diagram needs several diagrams apart from the ladder diagram itself given in figure 2. We list them in this appendix. In the various diagrams, the link joining vertices $i$ and $j$ is the Green function $G_{ij}$. All the vertices are integrated over the toroidal worldsheet. Diagrams can also involve derivatives of the Green function, and they are built using figure 4. 

\subsection{Diagram with two vertices}

\begin{figure}[ht]
\begin{center}
\[
\mbox{\begin{picture}(95,20)(0,0)
\includegraphics[scale=.45]{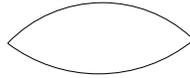}
\end{picture}}
\]
\caption{The diagram $\mathcal{T}_2$}
\end{center}
\end{figure}

The diagram with two vertices $\mathcal{T}_2$ is given in figure 24. It is defined by
\be \mathcal{T}_2 = \frac{1}{\tau_2^2} \int_{12} G_{12}^2. \ee

\subsection{Diagram with three vertices}

\begin{figure}[ht]
\begin{center}
\[
\mbox{\begin{picture}(95,50)(0,0)
\includegraphics[scale=.6]{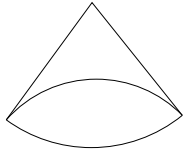}
\end{picture}}
\]
\caption{The diagram $\mathcal{T}_3$}
\end{center}
\end{figure}

The diagram with three vertices $\mathcal{T}_3$ is given in figure 25. It is defined by
\be \mathcal{T}_3 = \frac{1}{\tau_2^3} \int_{123} G_{12}^2 G_{13} G_{23}. \ee

\subsection{Diagrams with four vertices}

\begin{figure}[ht]
\begin{center}
\[
\mbox{\begin{picture}(225,75)(0,0)
\includegraphics[scale=.65]{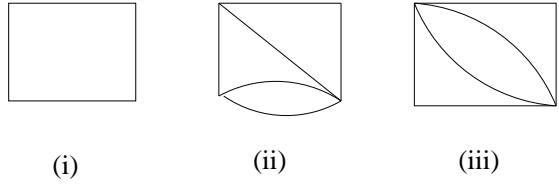}
\end{picture}}
\]
\caption{The diagrams (i) $\mathcal{F}_1$, (ii) $\mathcal{F}_2$ and (iii) $\mathcal{F}_3$}
\end{center}
\end{figure}

Apart from the ladder diagram $\mathcal{L}$ itself, the other diagrams with four vertices $\mathcal{F}_1$, $\mathcal{F}_2$ and $\mathcal{F}_3$ are given in figure 26. They are defined by
\bea \mathcal{F}_1 &=& \frac{1}{\tau_2^4} \int_{1234} G_{12} G_{23} G_{34} G_{41}, \non \\ \mathcal{F}_2 &=& \frac{1}{\tau_2^4} \int_{1234} G_{12} G_{23} G_{34} G_{41}^2 G_{24}, \non \\ \mathcal{F}_3 &=& \frac{1}{\tau_2^4} \int_{1234} G_{12} G_{23} G_{34} G_{41} G_{24}^2. \eea

\subsection{Diagrams with five vertices}

\begin{figure}[ht]
\begin{center}
\[
\mbox{\begin{picture}(275,80)(0,0)
\includegraphics[scale=.65]{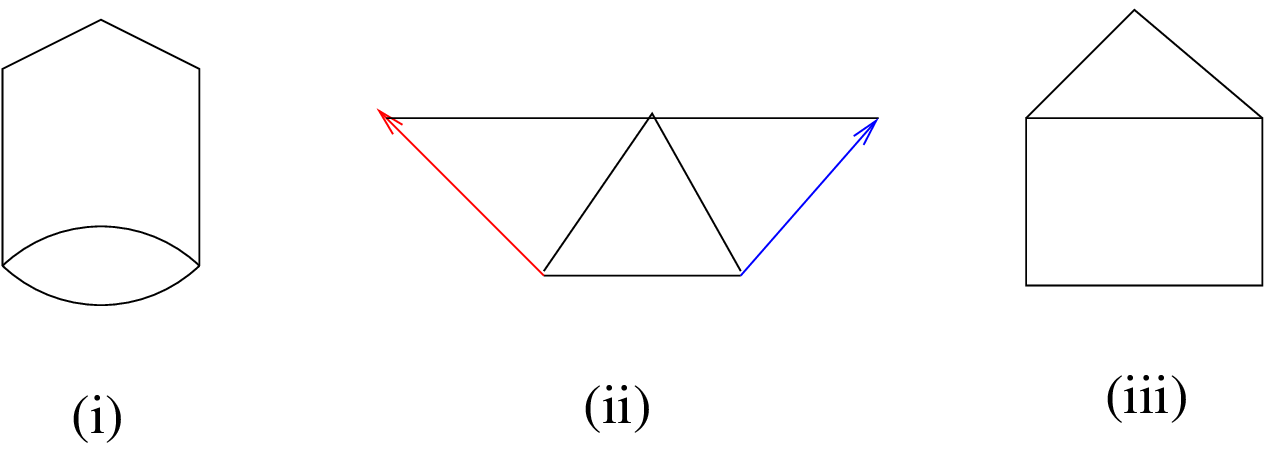}
\end{picture}}
\]
\caption{The diagrams (i) $\mathcal{V}_1$, (ii) $\mathcal{V}_2$ and (iii) $\mathcal{V}_3$}
\end{center}
\end{figure}

The diagrams with five vertices $\mathcal{V}_1$, $\mathcal{V}_2$ and $\mathcal{V}_3$ are given in figure 27. They are defined by
\bea \mathcal{V}_1 &=& \frac{1}{\tau_2^5} \int_{12345} G_{12} G_{23} G_{34} G_{45} G_{51}^2, \non \\ \mathcal{V}_2 &=& \frac{1}{\tau_2^5} \int_{12345} \p_2 G_{12} G_{23} G_{34} \bar\p_4 G_{45} G_{51} G_{13} G_{35}, \non \\ \mathcal{V}_3 &=& \frac{1}{\tau_2^5} \int_{12345} G_{12} G_{23} G_{34} G_{45} G_{51} G_{24}. \eea

We also consider the diagram $\mathcal{V}_4$ (this is called $D^{\wedge\p}_{11111}$ in~\cite{Green:2013bza}, upto an overall sign)
\be \mathcal{V}_4 = \frac{1}{\tau_2^5} \int_{12345}  G_{12} G_{23} G_{34} G_{45} G_{51} \p_3 G_{13} \bar\p_3 G_{35}\ee
which is needed in the main text. It is given in figure 28. 

\begin{figure}[ht]
\begin{center}
\[
\mbox{\begin{picture}(120,40)(0,0)
\includegraphics[scale=.55]{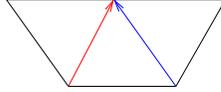}
\end{picture}}
\]
\caption{The diagram $\mathcal{V}_4$}
\end{center}
\end{figure}

\subsection{Diagram with six vertices}

\begin{figure}[ht]
\begin{center}
\[
\mbox{\begin{picture}(40,50)(0,0)
\includegraphics[scale=.4]{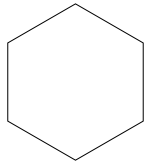}
\end{picture}}
\]
\caption{The diagram $\mathcal{S}_1$}
\end{center}
\end{figure}

The diagram with six vertices $\mathcal{S}_1$ is given in figure 29. It is defined by

\be \mathcal{S}_1 = \frac{1}{\tau_2^6} \int_{123456} G_{12} G_{23} G_{34} G_{45} G_{56} G_{61}.\ee


\providecommand{\href}[2]{#2}\begingroup\raggedright\endgroup

\end{document}